\documentclass[11pt,twoside]{article}
\usepackage{asp2010}

\resetcounters

\begin{document}
\newcommand{\msun}{\,\mbox{$\mbox{M}_{\odot}$}}

\title{Magnetic braking in convective stars}
\author{Gaitee A.J. Hussain
\affil{ESO, Karl-Schwarzschild-Str. 2, Garching bei M\"unchen, 85748 Germany}
}

\begin{abstract}

Magnetic braking causes the spin-down of single stars as they evolve
on the main sequence. Models of magnetic braking can also explain the evolution of 
close binary systems, including cataclysmic variables. 
The well-known period gap in the orbital period distribution of 
cataclysmic variable systems indicates that 
magnetic braking must be significantly disrupted in secondaries that are
fully convective. 
However, activity studies show that
 fully convective stars are some of the most active stars
observed in young open clusters.There is therefore conflicting evidence about what
happens to magnetic activity in  fully convective stars. 

Results from  spectro-polarimetric studies of cool stars have found that 
the field morphologies  and field strengths are dependent on  spectral type 
and rotation rate.  While rapidly rotating stars with radiative cores 
show strong, complex magnetic fields, they have  relatively weak dipole components.
Fully convective stars that are rapidly rotating also possess strong magnetic
 fields, but their configurations are much simpler; often close to dipole fields.
 
How this change in field geometry affects the stellar wind is the focus of 
several ongoing modelling efforts. Initial results suggest that rapidly rotating active dwarfs drive much stronger winds, about two orders of magnitude larger than
 those  on the Sun.
\end{abstract}

\section{Introduction}

The idea of stellar magnetic fields driving angular momentum
loss can be dated back to the 1962 paper by Evry Schatzman.
This paper brought together the ideas of the day to describe how
the  Hertzsprung-Russell diagram can be split into distinct sections. 
Stars with slow rotation rates are in the lower right part  of the diagram -- 
they are predominantly stars with outer convective envelopes.
As convective stars  host solar-type dynamo activity, 
stellar magnetic fields keep material in the extended magnetosphere
in corotation, thus exerting a braking torque and driving the spin-down
of a single star on the zero age main sequence. 

Magnetic braking should therefore operate in all systems with 
low mass stars ($0.4 < M_* < 1.5 $\msun); i.e., stars with outer convective envelopes.
Magnetic braking in the low mass secondary star of close binary systems is also 
responsible for determining 
binary separations. The secondary loses angular momentum and  
angular momentum is then removed from binary systems through tidal locking,
 causing the binary  separation to decrease and evolve further (Mestel 1968).

The first observational evidence for magnetic braking was gathered from a rotation study of 
G-dwarfs in open clusters of different ages (e.g., Pleiades, Hyades) and in the field. The seminal paper by
Skumanich (1972) found that these stars spin down following the inverse
square root of their age, $v \sim t^{1/2}$.
This braking law was adopted to derive magnetic braking laws for close binary stars with cool secondaries, assuming that secondary stars have a comparable mass loss rate to that of single G stars
(Verbunt  \& Zwaan 1981). However, it is worth noting that the Skumanich results
are based only on G stars with $v_e \sin i$ values up to 30\,km/s. Close binary stars can far exceed these velocities.

\begin{figure}
\plotone[height=6cm]{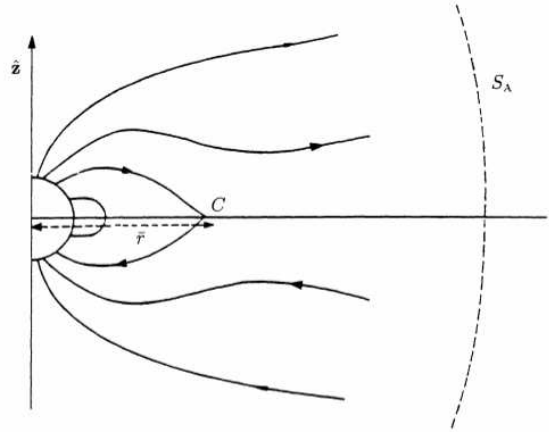}
\caption{ A schematic diagram of the wind model of Mestel and Spruit (1987).  
Here C marks a 
position inside the ``dead zone'' 
while the Alfv\'en surface, $S_A$, denotes the start of the ``wind zone''
where the field is radial.  This figure has
been reproduced from Campbell (1997) with permission.}
\end{figure}

This early work led to the development of a variety of angular momentum loss 
formulations. 
Iniitally braking laws were developed assuming symmetric winds flowing along purely radial field lines (Weber \& Davis 1967, Mestel 1968). 
Later formulations of braking laws allow for a more complex two-component coronal structure with an inner closed region and an outer region in which 
field lines are open (Figure 1; 
 Mestel 1984, Kawaler 1988, Mestel and Spruit 1987, Tout and Pringle 1992,  Ivanova and Taam 2003).
Solar eclipse images provide support for this large-scale coronal model as a first approximation. 

In these later models, the star has a large-scale dipole field
that  causes  a ``dead zone'' near the equator in which field lines are closed.
As matter is trapped in this zone, it cannot escape or 
contribute to the angular momentum loss and reduces
the efficiency of the magnetic braking.
The hot expanding corona is driven by thermal pressure gradients and 
centrifugal acceleration and causes the formation of 
an outer zone (the ``wind zone'') where the field is open. 

The field of the star gets  distorted where the kinetic energy density of the outflowing material matches the poloidal 
magnetic field density. Field lines are blown open into the flow where the poloidal velocity matches the Alfv\'en speed; where the kinetic energy density matches the poloidal magnetic energy density. This can be  calculated using the poloidal magnetic field, $B_{pol}$ and the mass density, $\rho$:

\begin{equation}
v_A=\frac{B_{pol}}{\sqrt{(\mu_{\circ}\rho)}}.
\end{equation}

Other variations are possible:
Tout and Pringle (1992) propose a stellar field that declines more strongly with distance due to more complex stellar fields; they  also assume that not all open field lines connect with the whole stellar surface.  Ivanova and Taam (2003) assume that the X-ray luminosity of the secondary star is generated in the dead zone and therefore use X-ray observations of stars to model the volume of the dead zone.  Just by this modification they find that their magnetic braking prescription can reproduce the observed rotation rates at a range of masses. 
It is clear that the field configuration is important in determining basic properties
of the dead zone and where the wind zone starts. 
X-ray measurements alone cannot reveal the distribution of the 
underlying field geometry.

There are numerous published angular momentum braking laws; these all rely on a series of assumptions
that have been treated differently. 
 Knigge et al. (2011, {\em this volume}) demonstrate most effectively
how these laws differ, and in fact can predict opposite 
trends with orbital period and stellar mass. We clearly need to understand more about the nature of stellar magnetic braking from observations of convective stars.

\section{Rotational evolution of stars: observations}

Stellar winds (or outflows) are very difficult to detect directly in main sequence cool stars. 
On the Sun the wind causes  a mass-loss rate
of {\em \.{M}}$=2\times 10^{-14}$ \msun\ yr$^{-1}$ (e.g., Feldman et al. 1977).
However, its low density and high temperature make it difficult to detect.
Direct measurements of outflows on other main sequence stars are even 
more challenging. 

Indirect measurements can be made through observations of 
the interaction between the stellar wind and the local interstellar medium.
This is  detected as extra Ly$\alpha$ absorption in  UV spectra from 
 the {\em Hubble Space Telescope}, HST (Wood et al. 2002, 2005).
 HST studies of a handful of systems reveal that mass loss rates should scale with
 magnetic activity.  However, the scaling relation depends on very few systems, some
 of which are binaries and therefore not well understood.

Studies of close binary systems and cool stars have been conducted to 
characterise mass loss and angular momentum loss rates further. These are described below. While  the techniques differ a coherent picture is starting to emerge.

\subsection{Magnetic braking in cataclysmic variable systems (CVs)}

Magnetic braking has been shown to explain the evolution of close binary systems.
All close binary systems lose  angular momentum through gravitational radiation. 
The  angular momentum loss rate due to gravitational radiation should 
depend on the mass of the component stars, $M_1$ and $M_2$ and the orbital separation
$a$, as follows 
(Paczy\'nski 1967, Knigge et al. 2011):
  \begin{equation}
dJ_{GR}/dt \sim \frac{M_1^2M_2^2M^{1/2}}{a^{7/2}}.
\end{equation}

Gravitational radiation clearly weakens in systems with large orbital separations.
The observed mass transfer rates in long period cataclysmic variable binary and low mass X-ray binary systems can only be explained  by another angular momentum loss mechanism  (Verbunt \& Zwaan 1981).
This is attributed to magnetic braking, with the assumption that convective stars in close binary systems will show the same magnetic braking levels as those seen in single convective stars. 
Their magnetic braking law prescription  depends on the secondary mass, $M_2$, radius, $R_2$ and rotation, $\Omega$ as shown below:
 
\begin{equation}  dJ_{MB}/dt \sim M_2R_2^4\omega^3.
 \end{equation}

The orbital period distribution of CVs is shown in  Figure 2 (Davis et al. 2008).
This has a largely bimodal 
distribution, with very few systems in the so-called ``gap'' between
orbital periods of 2-3 hours. In order to explain the accretion rates and sizes of the donor stars observed above the period gap, it is necessary to invoke magnetic braking.
At  orbital periods of 3 hours secondary stars become fully convective and presumably
this causes its magnetic activity to be disrupted and essentially switches off the 
magnetically driven wind (outflow).
As magnetic braking switches off, the donor shrinks within its Roche lobe, mass transfer is shut off and the secondary re-attains thermal equilibrium. The system is then no longer observed as a cataclysmic variable
and subsequent orbital evolution of the CV is driven by gravitational radiation only. Mass transfer resumes again when the secondary star makes contact with its Roche lobe once more at $P_{\rm orb} \sim$ 2h.

\begin{figure}
\plotone[height=6cm]{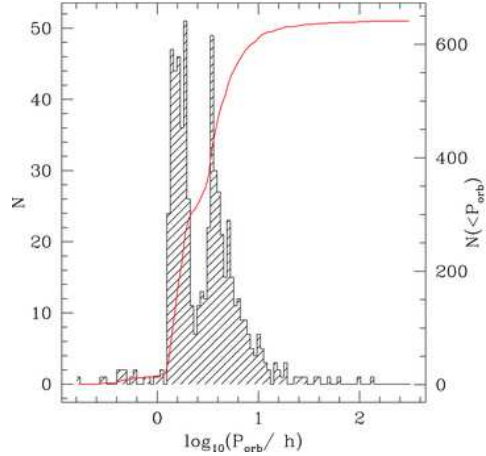}
\caption{ Orbital period distribution  of cataclysmic variable binary systems. There are few systems 
with periods  between 2-3 hours. This is attributed to a disruption in magnetic braking as stars
become fully convective. Reproduced from 
Davis et al. (2008) with permission.}
\end{figure}

An important proof of this scenario is provided in Patterson et al (2005).
They find that the masses of the donor stars above and below the period gap are very similar, which  is expected if mass transfer is reduced so the masses remain largely unchanged.  In line with the magnetic braking scenario, 
donor stars above the period gap 
are more inflated than those below the period gap. 
New work  measuring the sizes of
the donor stars further supports these findings: 
Knigge et al. (2011) fit the sizes of the secondaries using  a parametrised
version of the Verbunt \& Zwaan braking law (Rappaport et al. 1983). 
The find that 
 gravitational radiation losses alone are also not sufficient to account for  the star sizes
 below the period gap, with  twice the level of gravitational radiation-driven braking 
required to explain the evolution below the period gap.

\subsection{Magnetic braking in post common envelope binary systems (PCEBs)}
Studies of PCEBs can  shed light 
on the nature of magnetic braking. 
If magnetic braking  changes with secondary mass, 
the timescales for the onset of  accretion in close binaries should be affected.
Politano \& Weiler  (2006) use a Monte Carlo population synthesis code to 
investigate the relative distribution of PCEBS with 
low mass and high mass
secondaries assuming different magnetic braking formulations.

They find the most noticeable effect  when  magnetic braking
is completely disrupted in fully convective stars; this causes a significant decline 
in PCEB secondaries with radiative cores.
The relative number of PCEBs declines by 38\% in the mass bin at which 
 magnetic braking is switched on ($M_2>0.37$\msun).
Intermediate braking prescriptions are also investigated, in which the magnetic
braking is reduced in  rapidly rotating systems or the most X-ray active stars.
However, for both of the intermediate braking cases 
they find that the numbers of low mass and high mass secondaries remains similar. 
The fourth case investigated assumes  no magnetic braking, this finds a relative 
increase in the number of PCEBs with increasing secondary mass.

A large survey of white dwarf-main sequence (WDMS) binaries using the SDSS 
finds a decline of about 80\%
in the {\em relative fraction} of PCEBs at masses greater than $M_2>0.37$\msun\
(Schreiber et al. 2010). 
However, the simulation above predicts  the {\em relative number} of 
PCEBs not the fraction of PCEB/WDMS systems: the 
number of WDMS binary systems should increase 
with secondary mass. Taking this into account,
Politano \& Weiler's predictions  translate to  a decrease of  between 
38--73\% in the 
fraction of PCEB/WDMS systems at higher masses.
Schreiber et al. (2010) caution that, while the decrease they observe
 with increasing secondary star mass is in general agreement with the 
predictions from the disrupted magnetic braking model, the observed distribution is broader than that predicted. A contributing factor may be the uncertainty in the spectral type determination of these systems, 
which would effectively broaden an initially steep function. Alternatively, the onset of 
the disruption of magnetic braking  may occur more gradually rather than at one mass.

\begin{figure}
\plotone[width=7cm]
{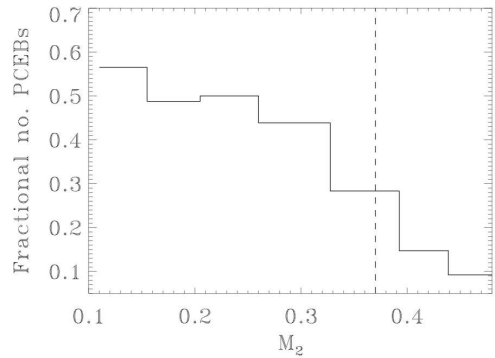}
\caption{
 The fractional number of PCEB systems ($n_{\rm PCEB}/n_{\rm WDMS}$)  relative to the total number of white dwarf main sequence star systems as a function  of secondary mass, $M_2$ (adapted from Schreiber et al. 2010). The dashed line denotes the fully convective boundary.} 
\end{figure}

\section{Magnetic activity and braking in single stars}
\label{sec:actconv}

Since Skumanich's 1972 study of G main sequence stars, we now have a wealth of information
tracking the angular momentum properties of convective stars over a range of spectral types (i.e. masses) and ages. Barnes (2003) collate rotation periods from open cluster
studies and find that the braking timescales 
depend strongly on a star's  spectral type, age, and crucially, the type
of magnetic activity behaviour displayed in that star.

He finds that stars ostensibly fall into two activity tracks: the more slowly rotating stars lie on the interface track, $I$: so-called as they show signs of classic interface (solar-type) dynamo activity.
They spin-down with age efficiently following a modified version of the Skumanich law. 
The other track is called the convective track, $C$: this tends to contain the more
 magnetically active, rapidly rotating stars and shows a reduced braking efficiency.
 Expressions governing the spin-down rates of single stars on these two tracks are shown below  (Barnes \& Kim 2010), where $k_C$ and $k_I$ are constants and $\tau_c$ is the convective turnover timescale:

\begin{equation}
\frac{dJ}{dt} = -  \Omega I_c k_C/\tau_c~; {\rm for\, the\, convective\, stars},
\end{equation}

\begin{equation}
\frac{dJ}{dt}= - \Omega^3 I_* 2\tau_c/k_I ~; {\rm for\, the\, interface\, stars}.
\end{equation}

 Stars move from the $C$ track to the $I$ track as they age, with lower mass stars taking 
 longer to make this transition. Almost all G stars will have made this transition by the first 200\,Myr on the main sequence, while M stars can take over 500\,Myr.

\section{Key questions}

Our current understanding of angular momentum evolution comes 
predominantly from statistical studies  of cool star systems. 
The root cause of the magnetic braking mechanism is of course
the stellar magnetic field. 
We can learn about the properties of stellar magnetic fields in 
more detail by studying proxies of magnetic 
activity such as X-ray emission. More recently,  
tomographic techniques have revealed even more detailed information
about the distribution and characteristics of magnetic fields  at the surfaces of stars
where they first emerge. 
In the rest of this paper we address
 the following questions. These have been posed by earlier studies
and serve to  place our understanding of the root causes of 
magnetic braking and stellar winds on a solid footing.

\begin{enumerate}

\item What is the dependence of stellar magnetic fields on spectral type, age, rotation and binarity? 
\item What happens to magnetic fields in fully convective stars?
\item Do magnetic fields look similar in single stars and in their binary star  counterparts? 
\item How does the stellar wind depend on the stellar magnetic field?

\end{enumerate}

\section{Magnetic fields in cool stars}

Magnetic braking laws that have been developed for  convective stars
rely on simple prescriptions for the stellar magnetic field, which has been 
variously modelled as a dipole or even as a monopole in some early cases. 

The magnetic activity state of a star is often characterised using 
a magnetic activity proxy, e.g., X-ray and Ca II H\&K emission.
These are measures of the magnetic heating in the outer atmospheres of cool stars
and, therefore, indirect measures of the magnetic flux threading through the stellar atmosphere.
Numerous X-ray studies of open clusters have revealed that 
X-ray luminosity and therefore magnetic activity levels are strongly dependent on rotation
rate, with X-ray emission generally increasing with increasing rotation.
A tighter correlation is found with the Rossby number, $R_{\circ}=P_{\rm rot}/\tau_c$; where $P_{\rm rot}=$ stellar rotation period and $\tau_c=$convective turnover time-scale.
$\tau_c$ is  a theoretical quantity that increases with increasing convection zone depth.

X-ray and Ca II activity studies have found that  the  
dynamo does {\em not} in fact switch off at full convection, as
predicted by studies of the evolution of close binary systems. 
Indeed, fully convective stars have similar fractional X-ray luminosities ($L_X/L_{\rm bol}$) to active  G and K stars (Figure 4;  Pizzolato et al.2003, James et al. 2000, Jeffries et al. 2011).  Furthermore,  there is still considerable 
uncertainty regarding exactly how these measures of magnetic heating relate
to the underlying magnetic field. While these diagnostics may be 
sensitive to the magnetic energy levels in stars,
they cannot reveal the underlying magnetic geometry and therefore the conditions that drive stellar winds. How magnetic flux is distributed in these stars is central to understanding  the conditions driving stellar winds. 

\begin{figure}
\plotone[width=12cm]{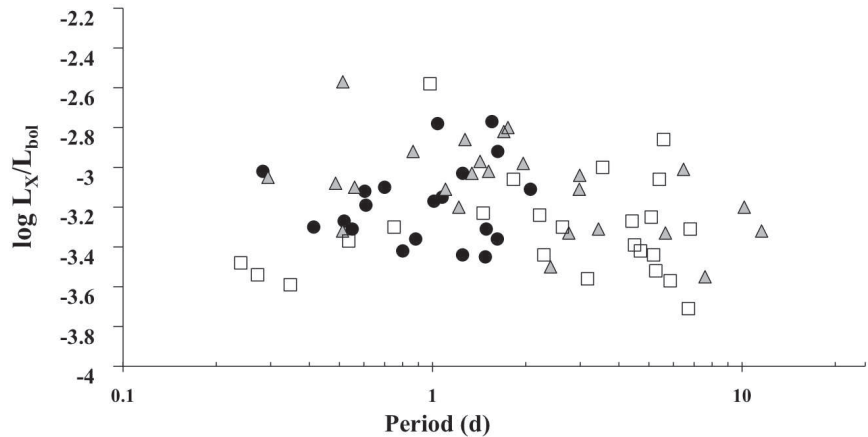}
\caption{Fractional X-ray luminosity as a function of rotation period for stars with different masses
($M_*<0.37$\msun: circles; $0.38<M_*<0.55$: grey triangles; $0.55<M_*<1.0$\msun: open squares).
The filled circles are fully convective systems and show no clear drop in X-ray luminosity (i.e. no drop in magnetic activity) Adapted from Jeffries et al. (2011).}
\end{figure}

\subsection{Spot maps}

The technique of Doppler imaging has been used to image the surfaces of over 80  
convective stars since it was first introduced in 1987 by Vogt, Penrod \& Hatzes 
(see review by Strassmeier 2009).
It can only be applied to rapidly rotating stars with $v_e \sin i > 15$km/s;
these stars are some of the most magnetically active, 
displaying X-ray luminosities up to two orders of magnitude greater 
than that on the Sun.
The dark starspots that are reconstructed 
are analogous to sunspots, which mark the largest
concentrations  of magnetic flux at the solar surface.

\begin{figure}
\plotone[height=8cm]{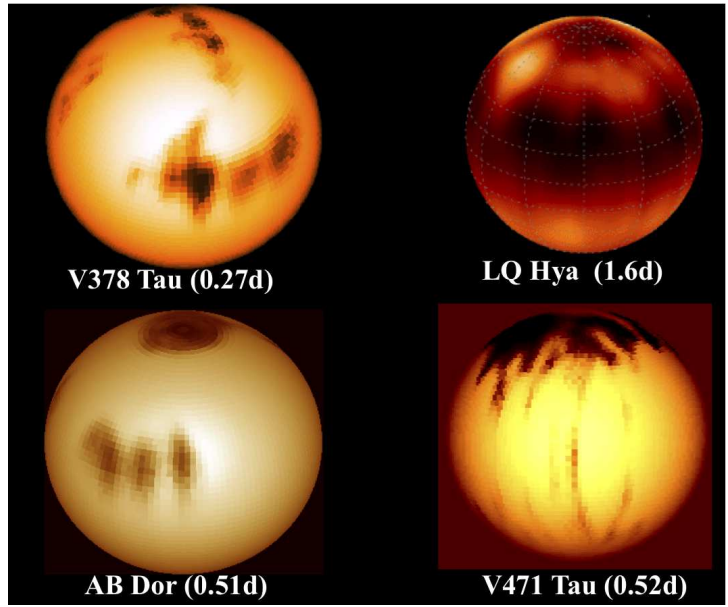}
\caption{Surface spots on four rapidly rotating K stars with similar magnetic activity levels (Hussain et al., 2000, 2006, Rice \& Strassmeier 1998). The rotation periods and names of all four stars are shown as captions. These images are snapshots of the stellar surface at a selected phase. }
\end{figure}

Surface spot maps show that stars with similar spectral types and rotation rates have 
similar spot patterns, suggesting that they not only have similar levels of activity but also 
similar magnetic flux emergence patterns. 
Figure 5 shows  Doppler maps from four K1-2 main sequence stars 
 with similar activity levels -- all of these maps show polar/high latitude spots 
 co-existing with  low latitude spots. 
Most G and K dwarfs tend to possess high latitude spots, with many showing large spots that cover their poles -- also known as polar
caps (e.g., Donati \& Collier Cameron 1997, Barnes et al. 2005, Jeffers et al. 2011).
G and K stars often have a mixture of both high latitude/polar spots and low latitude spots.
In early M dwarfs, that are not fully convective the starspot patterns change, with
little evidence for polar spots (Barnes et al. 2004).

The K2 secondary  in the post common envelope binary, V471 Tau,  is of particular interest.
It is instructive
to compare this map with those of single stars with  similar masses and rotation rates, 
such as AB Dor in Figure 5.
As V471 Tau has  an inclination angle of nearly 90$^{\circ}$
the low latitude spots cannot be reconstructed accurately and 
are smeared out due to a mirroring effect between the northern and southern hemispheres of the star. Despite this V471 Tau's spot maps  suggest that
there are shorter lived low latitude spots co-existing with the
 polar cap. This spot pattern is typical of other active K dwarfs and 
suggests that binarity and tidal locking do not fundamentally change the
magnetic field generation mechanism in stars  (Hussain et al. 2006). 
Our study of the tidally locked binary system, HD 155555 (G5 + K0), 
found spot and magnetic field patterns in the binary component stars
that are indistinguishable from those of single stars with similar spectral types
(Dunstone et al. 2008).

It is possible to map the surfaces of 
 the secondaries in CV systems using the technique of 
 Roche tomography (Watson \& Dhillon (2001), 
 which uses similar principles to those in Doppler imaging techniques. 
Maps of  AE Aqr, BV Cen and over eight other systems have now been published; they
show a mixture of high and low latitude spots though it is not clear
if these stars host polar spots due to the often low contrast reconstructions.
This is a particularly challenging technique as 
strong irradiation patterns across CV secondary surfaces can dilute
 the effects of starspots
 (e.g., QQ Vul; Watson et al.  2003).
 Furthermore, as CVs have short orbital periods and the secondaries are faint,
getting high S/N spectra in short exposure times (to limit phase smearing) limits the 
numbers of systems that can  be studied in this way with current facilities.

\subsection{Magnetic field maps}

With the advent of high resolution, high throughput spectro-polarimeters (e.g., CFHT-ESPADONS, 
ESO 3.6m-HARPSpol) we can now detect stellar magnetic fields directly using circular spectro-polarimetry.
Time-series of high resolution circularly polarised spectra are inverted to produce
surface magnetic field distributions 
using Zeeman Doppler imaging (Semel 1989, Donati \& Brown 1997). 
This technique applies Doppler imaging principles to 
high resolution circularly polarised profiles. 
As circularly polarised spectra are sensitive to the line-of-sight component of the 
stellar magnetic field, the technique enables us to measure the size of the magnetic field as well as reconstructing its geometry and distribution across the stellar surface.

Over 30 convective stars have been imaged using Zeeman Doppler imaging, covering a range of spectral types, rotation rates and evolutionary states. 
A summary of recent results  was presented in the review by Donati \& Landstreet (2009; see their Figure 3). Studies show  two clear transitions in magnetic activity characteristics.
The first happens with rotation rate: 
slowly rotating G and K-type  stars possess simple, mainly axisymmetric poloidal fields.
In more rapidly rotating stars, the field strengths increase and surface azimuthal  
fields -- horizontal, East-West oriented fields --  strengthen. The second transition in magnetic activity occurs in fully convective stars as described below.

\subsubsection{The transition to full convection}
\begin{figure}
\plotone[width=15cm]{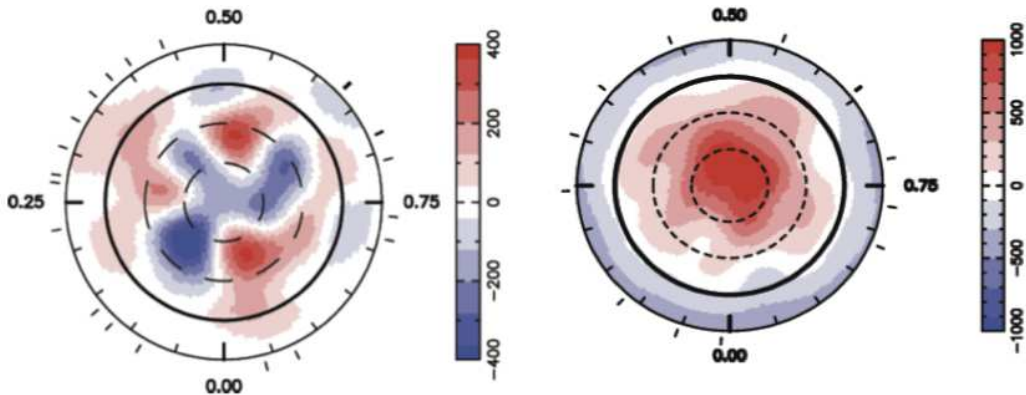}
\caption{Across the fully convective boundary: radial field maps of the surfaces of two M stars (Donati et al. 2008, Morin et al. 2008).
The colour scale represents magnetic flux in G. Left: DT Vir (M0.5, $P_{\rm rot}=2.9$\,d)
Right: EQ Peg B (M4.5, $P_{\rm rot}=0.4$\,d)}
\end{figure}

Magnetic field maps of M dwarfs show a marked change with mass. While
high mass M stars look similar to G and K-type stars, 
fully convective M stars ($M_* \leq 0.4$\msun) have predominantly poloidal field topologies that are more axisymmetric;
they also have stronger fluxes than their higher mass counterparts 
 (Figure 6; Donati et al. 2008, Morin et al. 2008).
This ties in with X-ray studies, which find no significant drop in X-ray luminosity of
 fully convective stars (Figure 4).

A more complete picture of magnetic field properties of convective stars is obtained  
by combining the field topologies from magnetic maps with measurements
of mean magnetic fluxes measured from intensity spectra. Because
circularly polarised spectra are only sensitive to the line-of-sight component of the
magnetic field, multiple switches in polarity across the surface
cause the circularly polarised signature to be diluted due to flux cancellation.
Measurements of Zeeman broadening in intensity profiles of magnetically  sensitive lines
make it possible to measure the mean magnetic flux at the surface of a star 
  (e.g., Johns-Krull \& Valenti 2000,  Reiners \& Basri 2009).

Reiners \& Basri (2009) compare their measurements of mean magnetic flux for stars covering a range of masses, 
$0.31\le M_* \le0.75$. Their results are summarised in Figure 7.
They find that the mean magnetic flux $<B_{\rm int}>$ is unaffected by the transition to 
full convection. However, the size of the magnetic flux recovered from 
from circularly polarised spectra ( $<B_{\rm circ}>$ in Figure 7) rises. This suggests that the field polarities
must be simpler and less prone to flux cancellation at lower masses, thus
supporting the results from Zeeman Doppler imaging studies. 

\begin{figure}
\plotone[width=14cm]{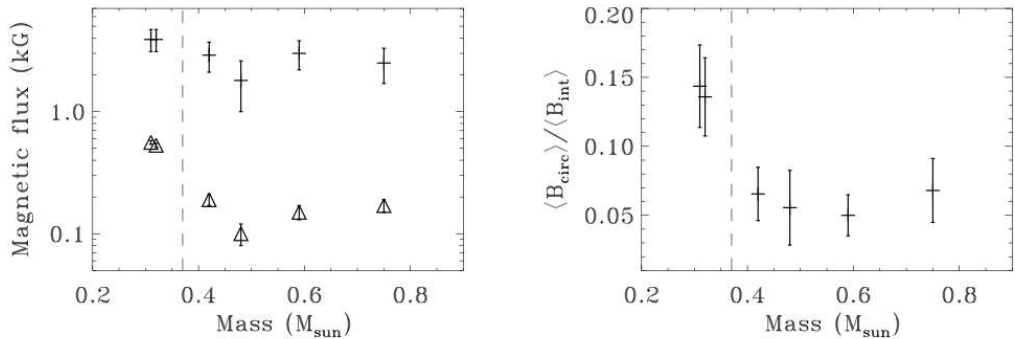}
\caption{Magnetic fields in M stars with different masses (Reiners \& Basri 2009, adapted with permission here).
{\em Left: } Mean magnetic fluxes recovered from intensity diagnostics (crosses) and circularly polarised profiles
(triangles). Circular polarisation recovers less flux overall than intensity; as circular polarisation is sensitive to the line-of-sight component of the stellar magnetic field, complex topologies will lead
to flux cancellation.
The mean magnetic fluxes from intensity, $<B_{\rm int}>$ (crosses),
 do not change significantly with stellar mass.
The dashed line denotes the fully convective boundary.
 {\em Right: } The fractional flux from circularly polarised profiles increases in fully convective stars,
 as  their  simpler magnetic field topologies result in reduced flux cancellation.}
 \end{figure}

\section{Wind models for cool stars}

Surface magnetic field maps can be extrapolated to produce detailed models
of a star's magnetosphere.  
The surface field maps are used to define the 
locations of footpoints of fields that extend in to the star's corona and beyond (Hussain et al. 2002).
A co-ordinated X-ray and Zeeman Doppler imaging study of  the K star, AB Dor, showed that the coronal model created by extrapolating the  surface
magnetic field map could reproduce the level of rotational modulation observed
in the contemporaneous X-ray 
 lightcurves and spectra  (Hussain et al. 2007).

Recent studies have shown how similar maps can be used to model stellar winds in detail using 
magnetohydrodynamic codes that were originally developed for the Sun 
 (Cohen et al. 2010, Vidotto et al. 2011).
These studies use the BATS-R-US code, and require the surface magnetic fluxes as an input, 
assuming a potential field initially.
Further inputs are the star's parameters and values for the 
base coronal density, $\rho_{\circ}$ and temperature, $T_{\circ}$.
The code assumes a thermal wind and allows the wind to evolve and interact with the magnetic field in a 
self-consistent way until a steady-state wind solution is reached.
The interaction between the coronal density structure and the speed of the wind determines 
 the angular momentum loss rate ({\em \.{J}}) and the mass loss rate ({\em \.{M}}) for the star.  

\begin{figure}
\plotone[width=15cm]{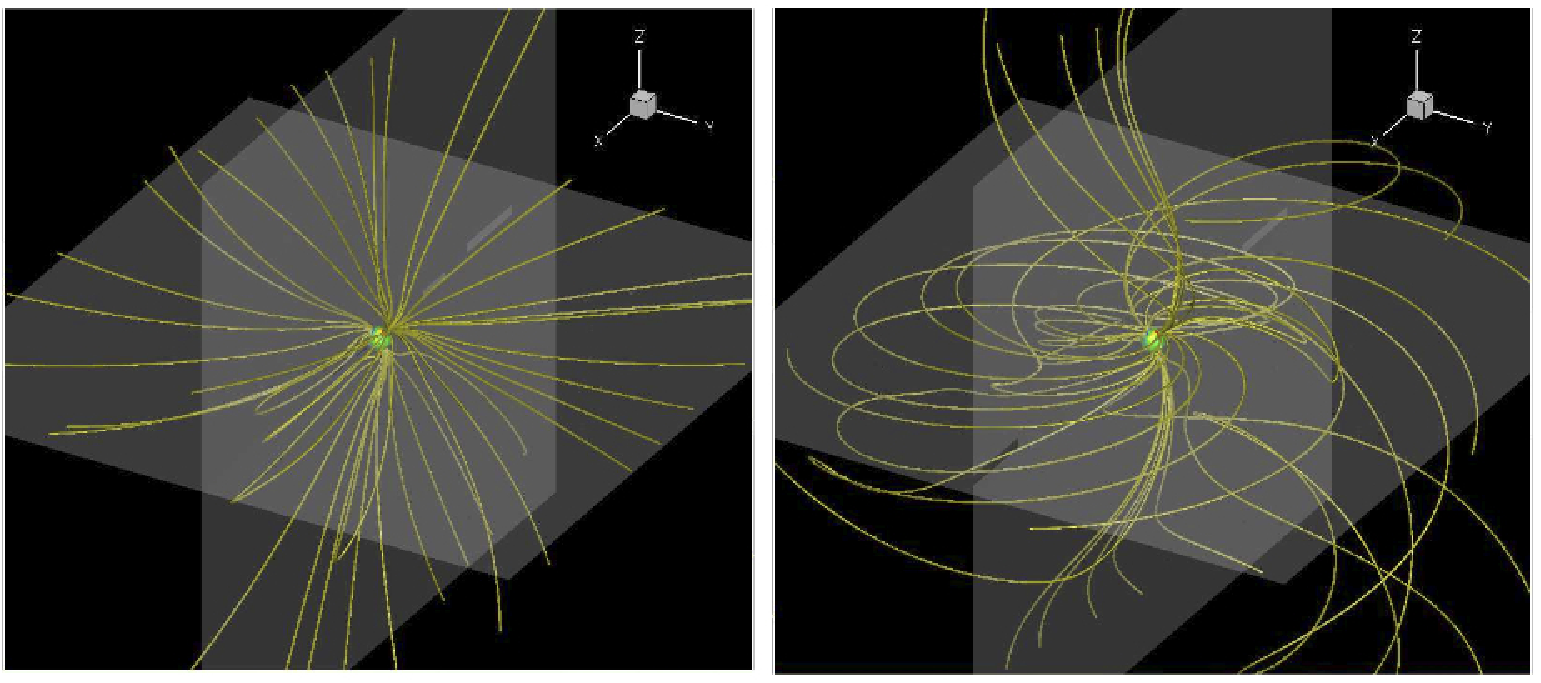}
\plotone[width=15cm]{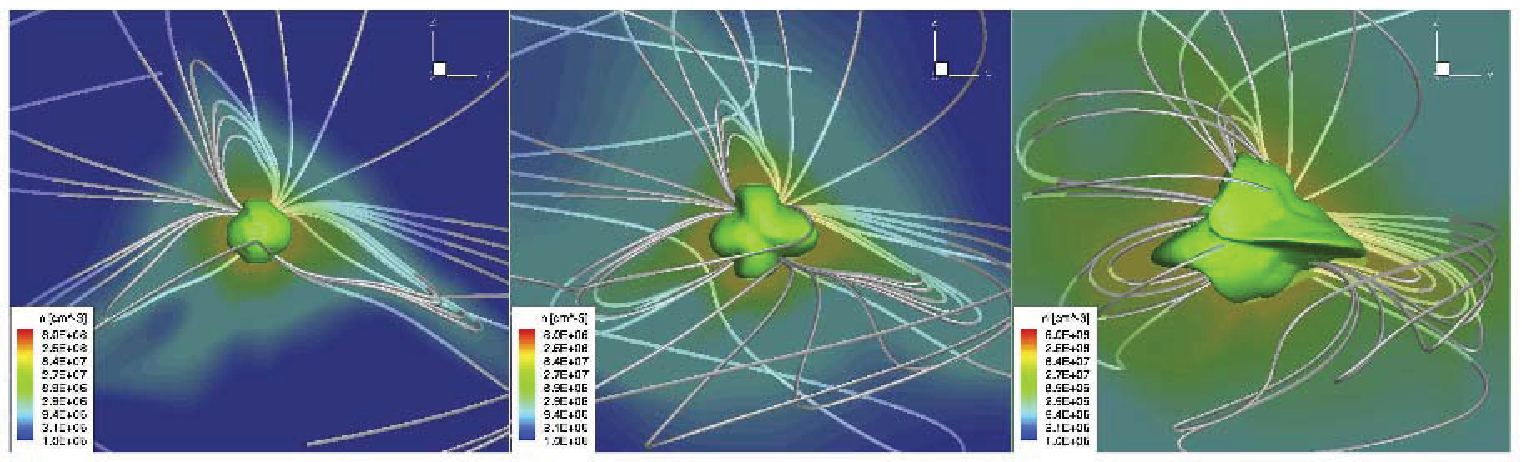}
\caption{{\em Top:} The effect of increasing $P_{\rm rot}$ on the stellar magnetosphere.
Spinning up AB Dor from $P_{\rm rot}=25$\,d (top left) to the actual
 $P_{\rm rot}=0.5$\,d results in greater field tangling.
{\em Bottom:} The effect of increased base coronal density from left to right, 
$n_{\circ}=2\times 10^8$, $10^{9}$ and $10^{10}$cm$^{-3}$. 
The iso-surface of $n=10^8$ cm$^{-3}$ is shown in green and the colour scale represents the density ($1E6$ to $8E8$cm$^{-3}$)
 (Cohen et al. 2010, with permission).}
\end{figure}

We find that wind models of the rapidly rotating K0 star,  AB Dor ($P_{\rm rot}=0.5$\,d, $T_{\rm eff}=5000$K),  
 indicate significantly higher mass loss rates than those seen on the present-day Sun  
 (Cohen et al. 2010). 
 AB Dor's surface maps show a complex field distribution, with field strengths of over
 700\,G,  much of which is  concentrated at higher latitudes than on the Sun.
These models require the base coronal density as an input. 
This is inherently uncertain even though it can be estimated from
X-ray coronal diagnostics, which are likely somewhat higher than the density at the base of the wind.  We use values ranging between $n_{\circ}=2 \times 10^8$ and $10^{10}$cm$^{-3}$ to investigate the dependence of the star on this parameter.\\
Our simulations show that two factors affect the loss rates of the stars  (Figure 8):\\
a)  Rotation rate --  increasing AB Dor's rotation period  by a factor of 50 
(from $P_{\rm rot}$=0.5\,d to 25\,d) reduces the loss rate by up to  
{\em \.{J}$_{0.5}/$\.{J}$_{25} \sim$70} and  
{\em \.{M}$_{0.5}/$\.{M}}$_{25} \sim$ 10.  AB Dor's maps
show strong high latitude flux near the pole of the star. 
This combined with rapid rotation  results in the tangling 
of  field lines and thus more of the corona is closed, leading 
to larger loss rates than in slower rotators with similar magnetic 
field distributions. If the star is spun down more of the field becomes 
open and the loss rates are reduced considerably. \\
b)  Coronal base density -- an increase of a factor of 50 ($n_0= 2 \cdot 10^{8}$ to
10$^{10}$cm$^{-3}$) increases the loss rates by over
an order of magnitude ({\em \.{J}$_{10^{10}}/$\.{J}$_{10^8} \sim$10} and \\
{\em \.{M}$_{10^{10}}/$\.{M}}$_{10^8} \sim$ 20).
This is because greater mass at the base increases the mass flux through the closed ``dead zone'';
this decreases the density gradient with height, which effectively leads
to a greater torque on the rotating star. \\

An analogous study of  the mid-M star, V374 Peg (M4, P{\rm rot}=0.44\,d), finds similarly
large loss rates (Vidotto et al. 2011). 
V374 Peg's surface field is quite similar to that of EQ Peg A (Figure 6):
it  has a simple, largely dipolar  field, with a strength of 1660\,G 
(compared to only 1-2\,G on the Sun).
While the mass and angular momentum loss rates are similar to those predicted for AB Dor, 
the reasons are different. 
The models for V374 Peg show much faster winds than on AB Dor, with 
relatively little field tangling despite the rapid rotation and presumably 
due to the simple dipolar field in V374 Peg.
As with the AB Dor study, Vidotto et al. also find that the braking rate strongly depends on the coronal base density, which is not well-defined.

These first detailed wind modelling studies pose some interesting questions as they find that  braking is more efficient in rapidly rotating stars, 
including the simpler M dwarf fields. 
So how do we explain the observations, which suggest that, in fully convective 
M stars, the braking must become less efficient? 
Future studies will reveal much more about how mass loss and angular momentum
loss rates change with magnetic topology.
The wind modelling techniques should first be fine-tuned against the few 
observable quantities such as
the mass loss measurements made by Wood et al. (2005) in a handful of systems.

\section{Summary}

This review provides an overview of angular momentum evolution in both binary
and single star systems. 
A wealth of observations suggest that magnetic braking is likely to operate at all
spectral types. However the form of the magnetic braking changes with spectral type, 
rotation rate and activity state.
The fundamental processes controlling the magnetic braking efficiency have, as yet,
to be firmly established. Surface imaging studies of activity in cool stars have revealed
a promising avenue and go some way towards answering the key questions we pose
in this paper.

\begin{enumerate}
\item What is the dependence of stellar magnetic fields on spectral type, age, and rotation? \\
Clear changes are seen with spectral type and rotation rate. Magnetic field maps of 
G-M-type stars confirm that the magnetic field topology in active rapidly rotating G and K-type stars 
differs compared  to less active slowly rotating counterparts.
Rapidly rotating stars show stronger, more complex, fields; with strong flux typically at high latitudes. Slowly rotating stars have more axisymmetric, simpler dipolar fields with
much weaker field strengths. 

\item What happens to magnetic fields in fully convective stars?\\
Magnetic field studies find a  transition  in M stars near the fully convective boundary ($M_* \le 0.4$\msun): they possess simpler more axisymmetric dipolar fields 
of the type seen in slowly rotating stars, but with  field strengths up to  three orders of magnitude larger. 
This transition agrees well with where the radiative core becomes negligibly small and convective turnover timescales are expected to increase (Donati et al. 2008).
These results are consistent  with X-ray and Ca II studies, which cannot detect changes in magnetic topology and find that the magnetic heating in the upper atmospheres of 
these stars  does not switch off  or decrease at full convection. 

How the change in field topology affects the properties of the stellar wind has yet to be established and is the
focus of intensive modelling efforts.
Early results from spectro-polarimetric 
studies of very low mass stars 
suggest that the field topologies change again below masses of 0.15\msun: the magnetic field becomes weaker and complex (Morin et al. 2008). However, this requires further investigation as it  is based on a small sample and one exception, WX UMa, has been found. 

\item Do magnetic fields look similar in single stars and in their binary star  counterparts? \\
Spot and magnetic field maps of the tidally locked main sequence secondaries in binary systems, 
V471 Tau and HD 155555, look similar
to those of single G and K-type stars,  with strong flux at both high and low latitudes. 
Images of secondaries in CVs also suggest a mix of high and low latitude spots. However, the tomography of CV secondaries is very challenging as the  spot signatures are difficult to resolve in contrast to irradiation patterns. The direct comparison with main sequence stars of similar masses has yet to be done systematically.

\item How does the stellar wind depend on the stellar magnetic field?\\
Initial results suggest that rapidly rotating active dwarfs drive stronger winds than those in more
slowly rotating systems. This ties in well with observed mass loss rates. However, 
further work is needed to establish how changes in coronal topology directly
affect the angular momentum and mass loss rates in stars.  
Inputs that are used in these models also should be refined: for example 
magnetic field maps can be enhanced to account for missing flux from dark polar spots; 
the base coronal densities also need to be constrained.

\end{enumerate}

Binary evolution studies have established that there is a 
significant disruption to magnetic braking when stars become fully convective.  
Ostensibly this transition is congruent with the point at which magnetic field maps 
show a switch from complex multipolar fields to simple dipolar fields.
How strongly this transition should affect magnetic braking has yet to be understood.
Rotation studies of single stars indicate that active stars spin down on slower timescales
than their inactive counterparts regardless of mass 
(Barnes 2003, Barnes \& Kim 2010); furthermore the braking observed in single stars is weaker than that
needed to explain the properties of CV secondaries above and below the period gap (Knigge et al. 2011).

Future avenues of investigation need to address the following points:
a) modelling braking in moderately active stars by extrapolating surface magnetic field maps as inputs and refining these models through comparison with the few measurements of loss rates in cool stars;
b) once outflows/winds are better understood in single stars models of these winds can be used to investigate the evolution of close binary systems in more detail;
c) establishing whether the spot patterns found in CV secondaries are analogous with
their counterparts in PCEBs and single stars.
Further spot and magnetic field maps from Roche tomography and Doppler imaging studies of close binary systems will prove invaluable to establish this latter point.

\acknowledgements 

Linda and the OC's are warmly thanked for the seamless organisation and putting together this enjoyable and stimulating meeting.

\bibliography{author}

\section{References}
  Barnes, J. R., Collier Cameron, A. , Lister, T. A., Pointer, G. R., Still, M. D.,   2005, MNRAS, 356, 1501 \\
  Barnes, J. R., James, D. J., Collier Cameron, A., 2004, MNRAS, 352, 589 \\
  Barnes, S.A., 2003, ApJ, 586, 464 \\ 
 Barnes, S.A., Kim, Y.-C., 2010, ApJ, 721, 675 \\ 
 Campbell, C.G., 1997, MNRAS, 291, 250 \\
 Cohen, O., Drake, J. J., Kashyap, V. L., Hussain, G. A. J., Gombosi, T. I., 2010, ApJ, 721, 80\\ 
 Davis, P. J., Kolb, U., Willems, B., G\"ansicke, B. T., 2008,   MNRAS, 389, 1563 \\ 
 Donati, J.-F., Collier Cameron, A.,     1997, MNRAS, 291, 1 \\
 Donati, J.-F., Brown, S.F., 1997, A\&A, 326, 1135 \\ 
 Donati, J.-F.,  Landstreet, J.D., 2009, ARA\&A, 47, 333\\ 
 Donati, J.-F., Morin, J., Petit, P., Delfosse, X., Forveille, T., Auri\`ere, M., Cabanac, R., Dintrans, B., Fares, R.,  Gastine, T., et al. 2008, MNRAS, 390, 545 \\ 
 Dunstone, N. J., Hussain, G. A. J., Collier Cameron, A., Marsden, S. C., Jardine, M., Stempels, H. C., Ramirez
 Velez, J. C., Donati, J.-F., 2008, MNRAS, 387, 481\\
Feldman, W. C., Asbridge, J. R., Bame, S. J., \& Gosling, J. T. 1977, in The Solar Output and its Variation, ed. O. R. White (Boulder: Colorado Associated Univ. Press), 351
 Hussain, G. A. J., Donati, J.-F., Collier Cameron, A., Barnes, J. R., 2000, MNRAS, 318, 961\\
 Hussain, G. A. J., van Ballegooijen, A. A., Jardine, M., Collier Cameron, A., 2002, ApJ, 575, 1078 \\ 
  Hussain, G. A. J., Allende Prieto, C., Saar, S. H., Still, M.,  2006, MNRAS, 367, 1699 \\ 
  Hussain, G. A. J., Jardine, M., Donati, J.-F., Brickhouse, N. S., Dunstone, N. J., Wood, K., Dupree, A. K., Collier Cameron, A., Favata, F. et al.  2007, MNRAS, 377, 1488\\ 
  Ivanova, N., Taam, Ronald E., 2003, ApJ, 599, 516\\ 
   James, D.J., Jardine, M.M., Jeffries, R.D., Randich, S., Collier Cameron, A., Ferreira, M., 2000, MNRAS, 318, 1217 \\ 
    Jeffers, S. V., Donati, J.-F., Alecian, E.,  Marsden, S. C., 2011, MNRAS, 411, 1301  \\
  Jeffries, R. D., Jackson, R. J., Briggs, K. R., Evans, P. A., Pye, J. P., 2011,
  MNRAS, 411, 2099\\ 
   Johns-Krull, C. M.; Valenti, J. A., 2000, {\em Stellar Clusters and Associations}, ASP Conf. Proc., Vol. 198. Eds., R. Pallavicini, G. Micela, and S. Sciortino, p.371 \\
 Kawaler, S.D., 1988, ApJ, 333, 236 \\ 
  Knigge, C., Baraffe, I., Patterson, J., 2011, ApJS, 194, 28 \\ 
  Mestel, L.,  1968, MNRAS, 138, 359 \\ 
 Mestel, L.,   1984, LNP, 193, 49 \\
  Mestel, L., Spruit, H.C., 1987, MNRAS, 226, 57 \\ 
  Morin, J., Donati, J.-F., Petit, P., Delfosse, X., Forveille, T., Albert, L., Auri\`ere, M., Cabanac, R., Dintrans, B., Fares, R., et al. 2008, MNRAS, 390, 567 \\ 
Paczy\'nski, B.,  1967, AcA, 17, 287 \\ 
  Patterson, J., Kemp, J., Harvey, D. A., Fried, R.E., Rea, R., Monard, B., Cook, L. M., Skillman, D.R., Vanmunster, T.,  Bolt, G.,  et al., 2005, PASP, 117, 1204 \\
  Pizzolato, N., Maggio, A., Micela, G., Sciortino, S., Ventura, P., 2003, A\&A, 
  397, 147 \\ 
  Politano, M., Weiler, K.P., 2006, ApJ, 641, 137 \\ 
    Rappaport, S., Verbunt, F., Joss, P.C., 1983, ApJ, 275, 713 \\
       Reiners, A., Basri, G., 2009, A\&A, 496, 787 \\ 
   Rice, J.B., Strassmeier, K.G., 1998, A\&A, 336, 972 \\
     Schatzman.  E., 1962, AnAp, 25, 18\\ 
 Schreiber, M.R., {G\"ansicke}, B.T., Rebassa-Mansergas, A., et al., 2010, A\&A, 513, 7 \\ 
 Skumanich, A.,1972, ApJ, 171, 565 \\ 
   Semel, M., 1989, A\&A, 225, 456\\ 
  Strassmeier, K.G.,  2009, A\&ARv, 17, 251  \\
  Tout, C.A., Pringle, J.E., 1992, MNRAS, 256, 269 \\  
Verbunt, F.,  Zwaan, C.,  1981, A\&A, 100, 7\\
Vidotto, A., Jardine, M., Opher, M., Donati, J.F., Gombosi, T.I., 2011, MNRAS, 412, 351\\
  Vogt, S.S., Penrod, G. D., Hatzes, A. P., 1987, ApJ, 321, 496\\
 Watson, C. A., Dhillon, V. S., 2001, MNRAS, 326, 67\\
 Watson, C.A., Dhillon, V. S., Rutten, R., Schwope, A.D., 2003, MNRAS, 341, 129\\
Weber, E.J., Davis, L. Jr,  1967, ApJ, 148, 217\\
Wood, Brian E., M\"uller, Hans-Reinhard, Zank, Gary P., Linsky, Jeffrey L., 2002, ApJ, 574, 412 \\
Wood, B. E., M\"uller, H.-R., Zank, G. P., Linsky, J. L., Redfield, S., 2005,  ApJ, 628, L143

\end{document}